\documentclass[11pt]{article}
\usepackage{moriond}

\def\be{\begin{equation}}
\def\ee{\end{equation}}
\def\bea{\begin{eqnarray}}
\def\eea{\end{eqnarray}}

\newcommand{\Photo}{}

\begin{document}
\vspace*{4cm}
\title{2HDM CONFRONTING LHC DATA}

\author{P.M.~FERREIRA}
\address{Instituto Superior de Engenharia de Lisboa - ISEL,
	1959-007 Lisboa, Portugal and Centro de F\'{\i}sica Te\'{o}rica e Computacional,
    Faculdade de Ci\^{e}ncias,
    Universidade de Lisboa,
    Av.\ Prof.\ Gama Pinto 2,
    1649-003 Lisboa, Portugal}
\author{RUI SANTOS}
\address{Instituto Superior de Engenharia de Lisboa - ISEL,
	1959-007 Lisboa, Portugal and Centro de F\'{\i}sica Te\'{o}rica e Computacional,
    Faculdade de Ci\^{e}ncias,
    Universidade de Lisboa,
    Av.\ Prof.\ Gama Pinto 2,
    1649-003 Lisboa, Portugal}
\author{MARC SHER}
\address{ High Energy Theory Group, College of William and Mary,  Williamsburg, Virginia 23187, U.S.A.}
\author{JO\~{A}O P.~SILVA}
\address{Instituto Superior de Engenharia de Lisboa,
	1959-007 Lisboa, Portugal and Centro de F\'{\i}sica Te\'{o}rica de Part\'{\i}culas (CFTP),
    Instituto Superior T\'{e}cnico, Universidade T\'{e}cnica de Lisboa,
    1049-001 Lisboa, Portugal}

\maketitle\abstracts{
Almost all data collected at the LHC during the 7 and 8 TeV runs has now been analysed by the ATLAS and CMS collaborations. 
Its consistency with the Standard Model (SM) predictions has cornered the CP-conserving two-Higgs doublet model (2HDM) into the
SM limit, $\sin (\beta -\alpha) = 1$. However, there are still allowed regions of the 2HDM parameter space away from this limit. In this work
we discuss how the 2HDM is performing in view of the LHC data together with the remaining available experimental and theoretical
constraints.}

\section{Introduction}

The discovery of a neutral Higgs boson~\cite{ATLASHiggs} at CERN's Large Hadron Collider (LHC)
has now been confirmed. Hence, all extensions of the Standard Model (SM) have to abide to the existence of a Higgs
boson with SM like properties and with a mass of around 125 GeV. The two-Higgs doublet model (2HDM)
is one of the simplest extension of the SM. It is built with the addition of a second complex scalar doublet to
the SM field content. Consequently, one obtains a richer particle spectrum with one charged and three neutral
scalars.  All neutral scalars could in principle be the scalar discovered at the LHC~\cite{Ferreira:2011aa}. 
However, a pure pseudo-scalar state with a 125 GeV mass has now been experimentally ruled out~\cite{CPnotes}. 
In the next section we present the 2HDM model and in the following sections we discuss how well
the most common CP-conserving 2HDM can accommodate the LHC data.

\section{The Model}
\label{sec:models}

The addition of a second complex scalar doublet to the SM's Yukawa Lagrangian 
gives rise to the appearance of scalar exchange
flavour changing neutral currents (FCNCs)
which are known to be strongly constrained by experiment.
Avoiding those dangerous FCNCs can be accomplished by simply 
imposing a $Z_2$ symmetry on the
scalar doublets, $\Phi_1 \rightarrow \Phi_1$,
$\Phi_2 \rightarrow - \Phi_2$,  and a corresponding symmetry to the
fermion fields. This leads to the well known four Yukawa model types I, II, Y (III, Flipped)
and X (IV, Lepton Specific)~\cite{barger}. 
%
The scalar potential in a softly broken $Z_2$ symmetric 2HDM can be written as
\begin{eqnarray}
V(\Phi_1,\Phi_2) & = & m^2_1 \Phi^{\dagger}_1\Phi_1+m^2_2
\Phi^{\dagger}_2\Phi_2 + (m^2_{12} \Phi^{\dagger}_1\Phi_2+{\mathrm{h.c.}
}) +\frac{1}{2} \lambda_1 (\Phi^{\dagger}_1\Phi_1)^2 +\frac{1}{2}
\lambda_2 (\Phi^{\dagger}_2\Phi_2)^2\nonumber \\ 
& &+ \lambda_3
(\Phi^{\dagger}_1\Phi_1)(\Phi^{\dagger}_2\Phi_2) + \lambda_4
(\Phi^{\dagger}_1\Phi_2)(\Phi^{\dagger}_2\Phi_1) + \frac{1}{2}
\lambda_5[(\Phi^{\dagger}_1\Phi_2)^2+{\mathrm{h.c.}}] ~, \label{higgspot}
\end{eqnarray}
where $\Phi_i$, $i=1,2$ are complex SU(2) doublets. All parameters except for 
$m_{12}^2$ and $\lambda_5$ are real as a consequence of the hermiticity
of the potential. We will focus on the
CP-conserving potential where $m_{12}^2$, $\lambda_5$ and the VEVs are all real.
In this model the three CP-eigenstates are usually denoted
by $h$ and $H$ (CP-even) and $A$ (CP-odd). 
As shown in~\cite{vacstab1}, once a CP-conserving 
vacuum configuration is chosen, all charge breaking stationary points are saddle
points with higher energy.  Hence, the 2HDM is stable at tree-level against charge 
breaking and once a non-charge breaking vacuum is chosen the model 
has two charged Higgs bosons that
complete the 2HDM particle spectrum.
We choose as free parameters of the model, the four masses, the rotation
angle in the CP-even sector, $\alpha$, the ratio of the vacuum expectation
values,  $\tan\beta=v_2/v_1$, and the soft breaking parameter redefined as $M^2=m_{12}^2/(\sin \beta \, \cos \beta)$ 
(see~\cite{Branco:2011iw} for a review of the model).
The remaining theoretical and experimental constraints used in this work were 
recently discussed in~\cite{Barroso:2013zxa}. 

\section{Results and discussion}

\begin{figure}[h!]
\centering
\includegraphics[width=3.0in,angle=0]{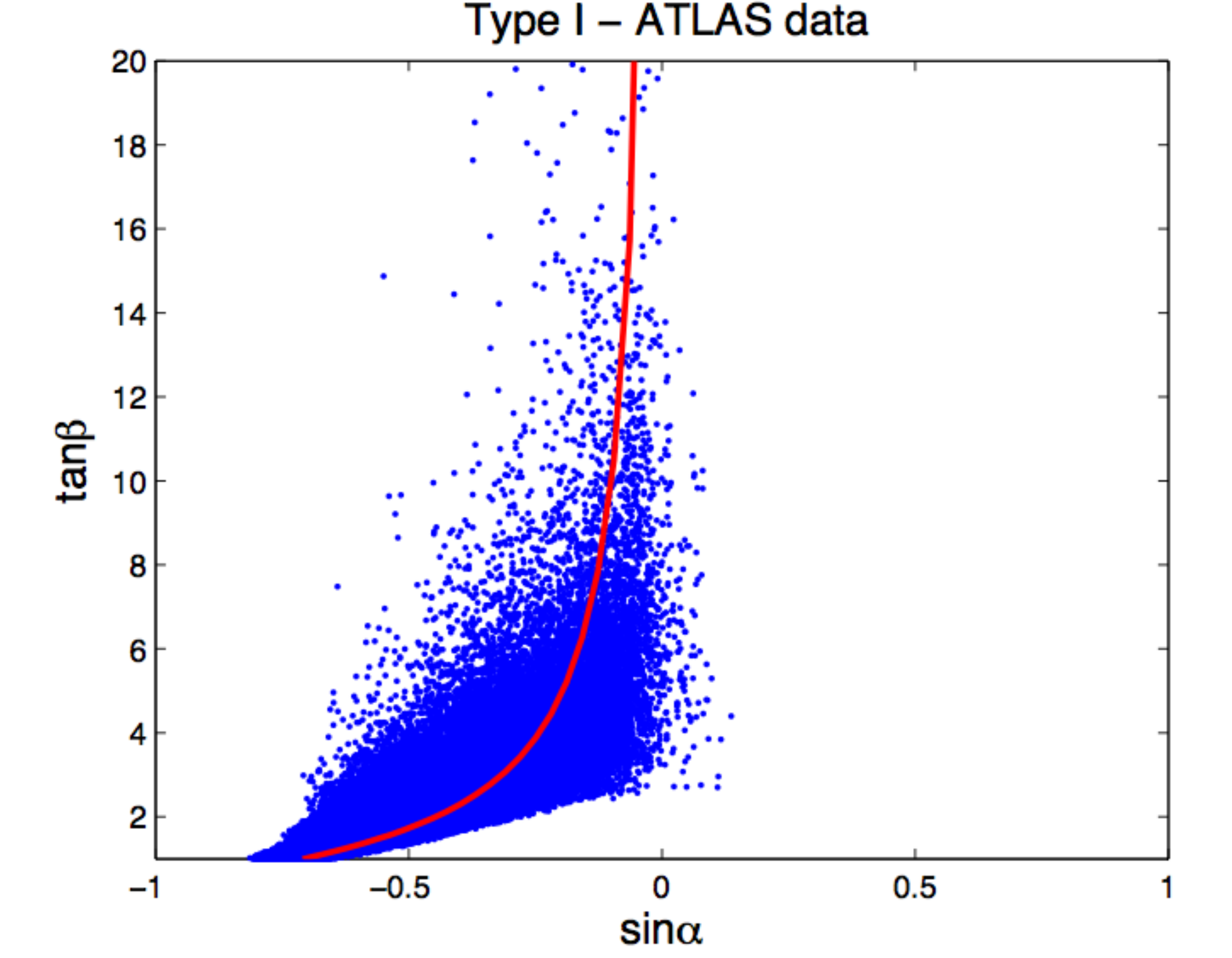}
\includegraphics[width=3.2in,angle=0]{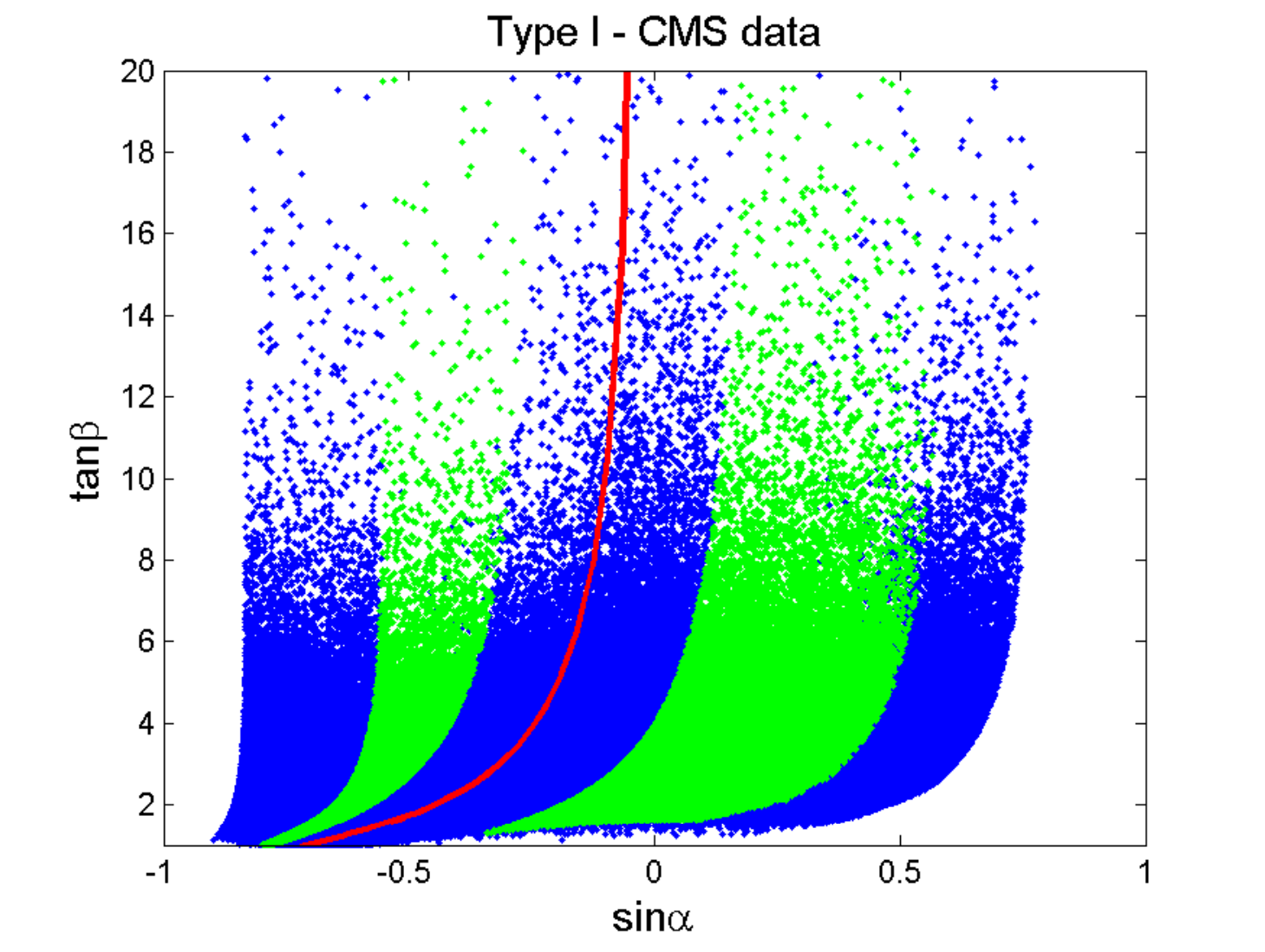}
\vspace{-0.3cm}
\caption{Points in the  ($\sin \alpha$, $\tan \beta$)  plane that passed all
 the constraints in model type I using the ATLAS data analysis (left) and using the CMS data analysis (right) at 1$\sigma$ in green (light grey) and 2$\sigma$ in blue (dark grey). 
Also shown is the
line for the SM limit $\sin(\beta - \alpha) =1$.}
\label{fig1}
\end{figure}

We randomly generate points in the model's
parameter space such that $m_h = 125$ GeV, 90 GeV   $ \leq m_A \leq$ 900 GeV,
$m_h < m_H \leq$ 900 GeV, $1 \leq \tan \beta \leq 40$, $- (900)^2$ GeV$^2$ 
$ \leq m_{12}^2 \leq 900^2$ GeV$^2$ and $-\pi/2 \leq \alpha \leq \pi/2$.
In order to respect the flavour constraints we take 90 GeV $ \leq m_{H^\pm} \leq$ 900 GeV for 
type I while
for type II the allowed range is 360 GeV $ \leq m_{H^\pm} \leq$ 900 GeV. 
We define the quantities $R_f$ as the ratio of the number of events
predicted in the 2HDM to that obtained in the SM for a given final state $f$.
\begin{equation} \label{Sratio}
R_f =
\frac{\sigma(pp \to h)_{\textrm{2HDM}}\
\textrm{BR}(h \to f)_{\textrm{2HDM}}}{
\sigma(pp \to h_{\textrm{SM}})\ \textrm{BR}(h_{\textrm{SM}} \to f)},
\end{equation}
where $h$ is the lightest CP-even Higgs (125 GeV), $\sigma$ is the Higgs production
cross section, BR the branching ratio,
and $h_{\textrm{SM}}$ is the SM Higgs boson.
In our analysis,
we include all Higgs production mechanisms,
namely,
gluon-gluon fusion using HIGLU at NLO~\cite{Spira:1995mt},
vector boson fusion (VBF)~\cite{LHCHiggs}, Higgs production in
association with either $W$, $Z$  or
$t\bar{t}$~\cite{LHCHiggs},
and  $b \bar{b}$ fusion~\cite{Harlander:2003ai}. A number of similar studies
where LHC data was used to constrain the 2HDM were presented in~\cite{many}.


\begin{figure}[h!]
\centering
\includegraphics[width=3.1in,angle=0]{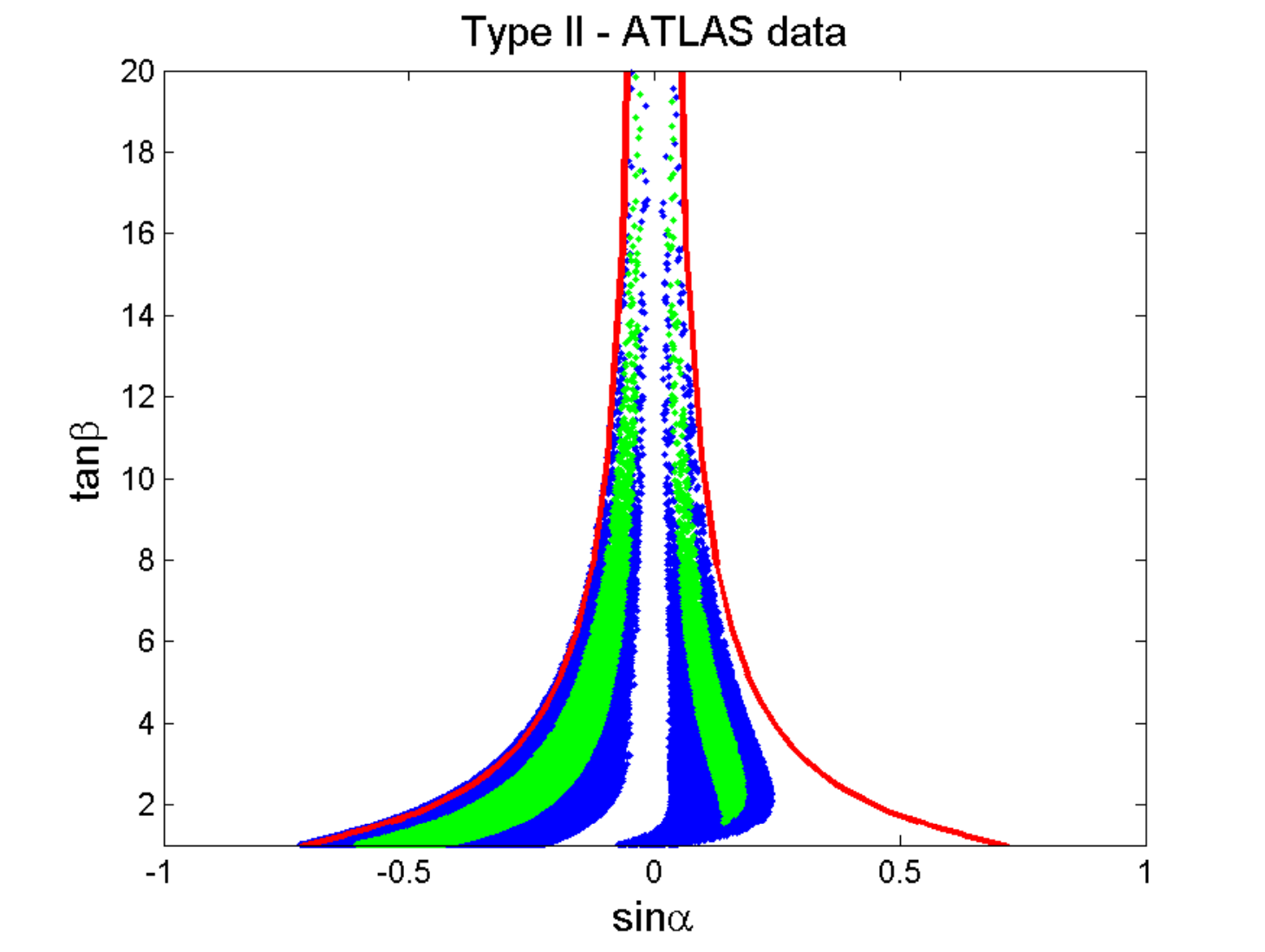}
\includegraphics[width=3.1in,angle=0]{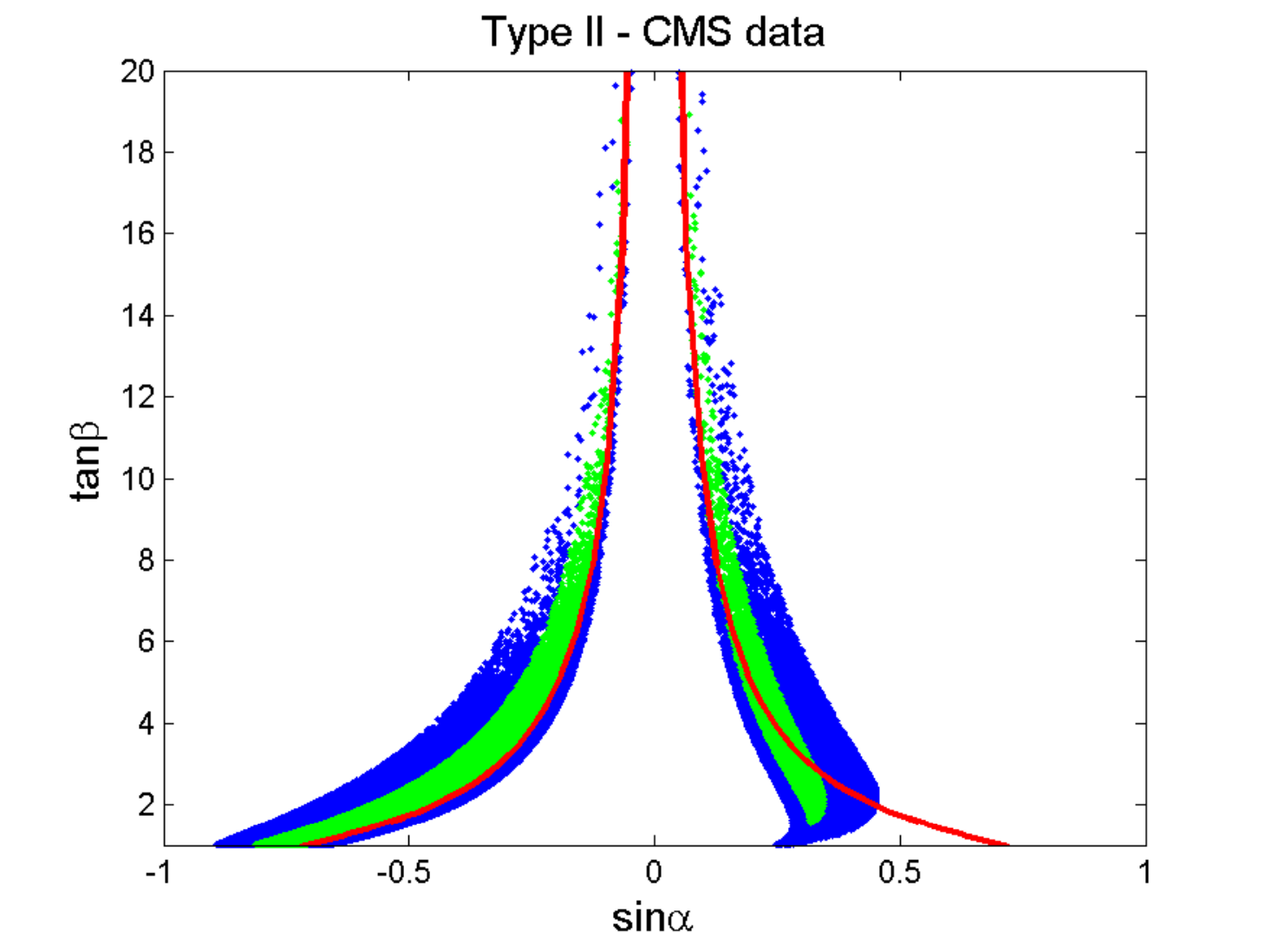}
\vspace{-0.3cm}
\caption{Points in the  ($\sin \alpha$, $\tan \beta$)  plane that passed all
 the constraints in model type II using the ATLAS data analysis (left) and using the CMS data analysis (right) at 1$\sigma$ in green (light grey) and 2$\sigma$ in blue (dark grey). 
 Also shown are the
lines for the SM limit $\sin(\beta - \alpha) =1$ (negative $\sin \alpha$) and for the limit
$\sin(\beta + \alpha) =1$.}
\label{fig2}
\end{figure}
\begin{figure}[h!]
\centering
\includegraphics[width=3.1in,angle=0]{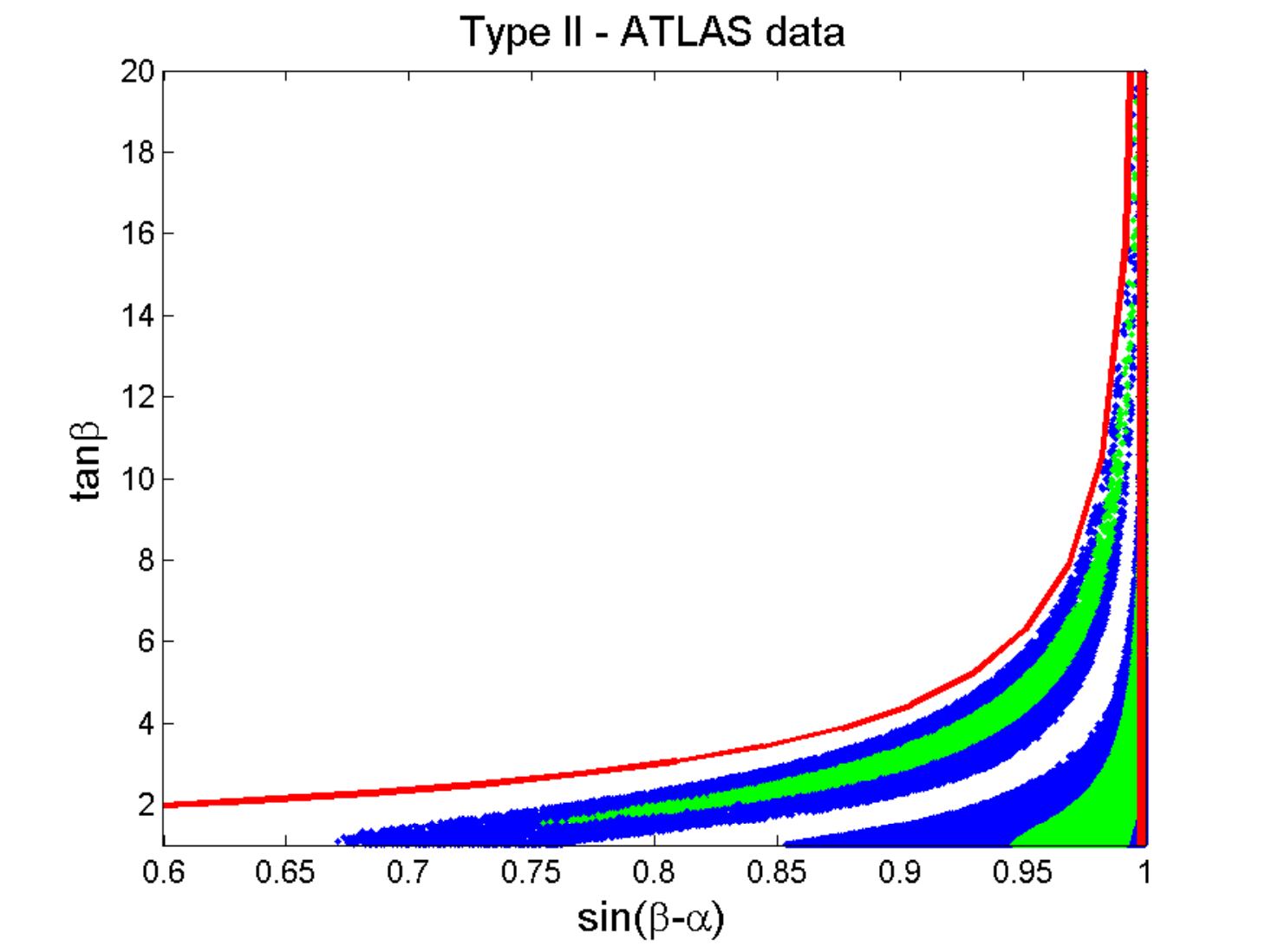}
\includegraphics[width=3.1in,angle=0]{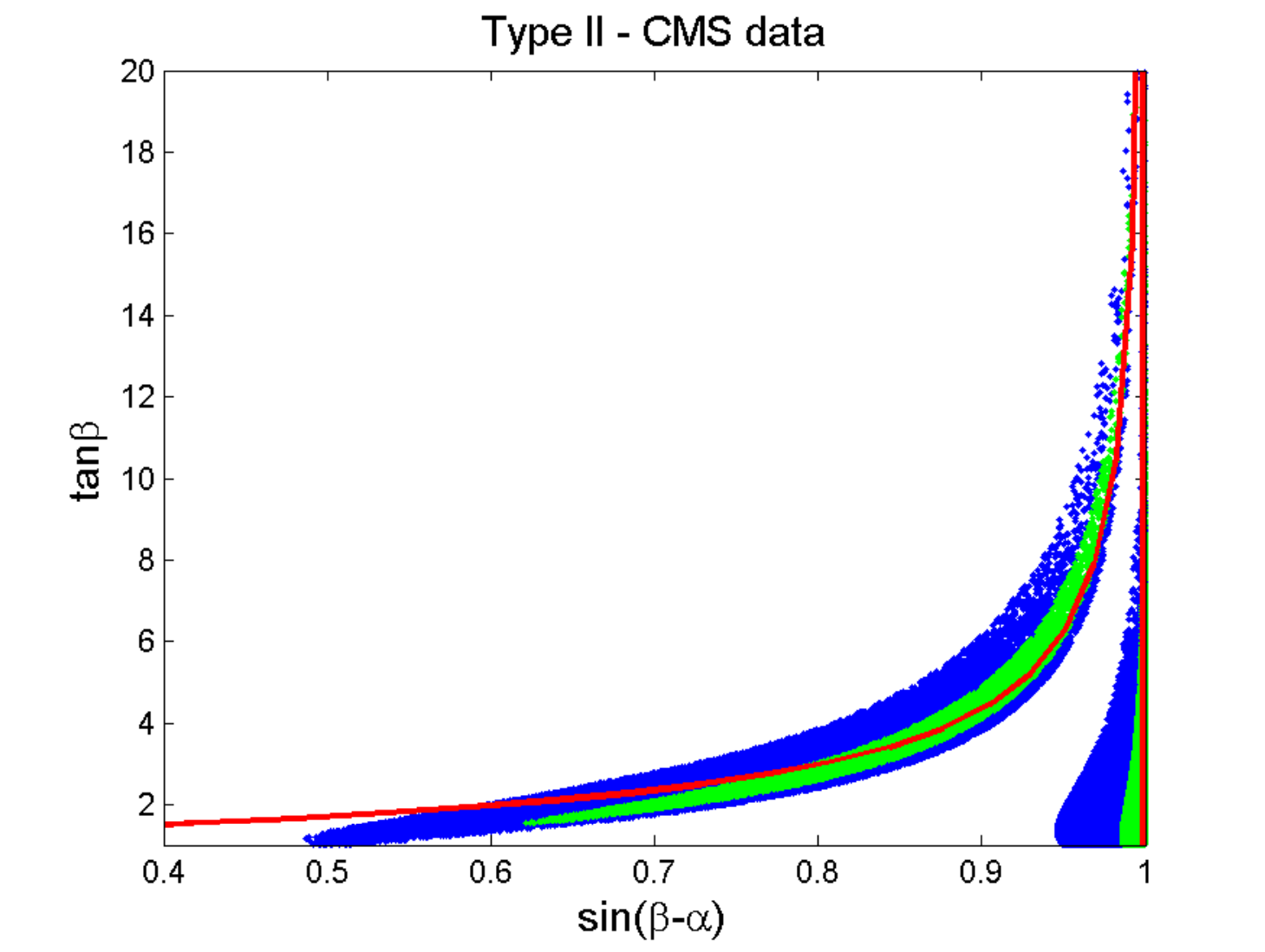}
\vspace{-0.4cm}
\caption{Points in the  ($\sin (\beta - \alpha)$, $\tan \beta$)  plane that passed all
 the constraints in model type II using the ATLAS data analysis (left) and using the CMS data analysis (right) at 1$\sigma$ in green (light grey) and 2$\sigma$ in blue (dark grey). 
 Also shown are the
lines for the SM limit $\sin (\beta - \alpha) =1$ and for the limit
$\sin(\beta + \alpha) =1$.}
\label{fig3}
\end{figure}

In figure~\ref{fig1} we present the results for the type I model  in the ($\sin \alpha$, $\tan \beta$)  plane
using the ATLAS data~\cite{ATLASnotes}  (left) and the CMS data~\cite{CMSnotes} (right).
The differences between the two plots are easy to understand. The ATLAS data analysis forces $R_{ZZ}$ to be large
but as we have previously shown~\cite{Barroso:2013zxa} $R_{ZZ}$ can never be above one in type I. Consequently,
no points survive at 1$\sigma$. With the CMS data analysis plenty of 1$\sigma$ points survive because all $R_{VV}$ are below one.
The 1$\sigma$ region  is slightly away from the SM limit because the central values of $R_{VV}$ are below one.
Further, we see in both plots a large dispersion of values around the SM limit. The reason is that 
$R_{VV} \approx \sin^2 (\beta -\alpha)$ in the limit where $BR(h \to b \bar b)  \approx 1$ and this function
is very sensitive from deviations from 1.

In figure~\ref{fig2} we now show similar plots for the type II model. In type II, $R_{VV}$ is not a sensitive quantity~\cite{Barroso:2013zxa}. Therefore
both the 1$\sigma$ and the 2$\sigma$ points are very close to the two limiting lines. The difference between the two plots
is that the points tend to concentrate below the limiting lines for ATLAS and above the same lines for CMS. This is of course
a consequence of the central values of the ATLAS $R_{VV}$ being above the SM while the CMS ones are below the SM expectation. 
Finally we present in figure~\ref{fig3} plots similar to the ones in figure~\ref{fig2} but now in the ($\sin (\beta - \alpha)$, $\tan \beta$)
plane. Again we see the same trend with the allowed points on opposite sides of the limiting lines, $\sin (\beta - \alpha) =1$ and
$\sin(\beta + \alpha) =1$. We should point out that remarkably (even being very conservative),
we can already say that values of $\sin (\beta - \alpha) $ below 0.5 are excluded at 2$\sigma$. Even more interesting is that 
for values of $\sin (\beta - \alpha) $ below say 0.8 $\tan \beta$ has to be below 4. Hence, in
type II large values of $\tan \beta$ are only allowed close to the SM limit. Note that a scan is performed varying
all parameters except $m_h = 125$ GeV with no need to define any benchmarks. Finally it is also interesting to note
that also the value of $\cos (\beta - \alpha)$ (and consequently the coupling of the heavy Higgs
to the gauge bosons) is already constrained by the light Higgs data~\cite{Barroso:2013zxa}.
\vspace{-0.4cm}
\section*{Acknowledgments}
The works of P.M.F. and R.S. are supported in part by the Portuguese
\textit{Funda\c{c}\~{a}o para a Ci\^{e}ncia e a Tecnologia} (FCT)
under contract PTDC/FIS/117951/2010, by FP7 Reintegration Grant, number PERG08-GA-2010-277025,
and by PEst-OE/FIS/UI0618/2011.
The work of J.P.S. is funded by FCT through the projects
CERN/FP/109305/2009 and  U777-Plurianual,
and by the EU RTN project Marie Curie: PITN-GA-2009-237920.  The work of MS is supported by the NSF under Grant No.~PHY-1068008.

\end{document}